\documentclass[%
reprint,
superscriptaddress,
amsmath,amssymb,
aps,
pre,
]{revtex4-1}

\usepackage{graphicx}
\usepackage{float}
\usepackage[export]{adjustbox}
\usepackage[caption=false]{subfig}
\usepackage{dcolumn}
\usepackage{bm}
\usepackage{mathtools}

\usepackage[list-final-separator = {{, }},range-units=single,list-units=single,range-phrase=-]{siunitx}

\DeclareSIUnit\molar{\textsc{M}}
\DeclareSIUnit\ec{\si{\elementarycharge}}

\newcommand{\be}{\begin{equation}}
\newcommand{\ee}{\end{equation}}
\newcommand{\bea}{\begin{eqnarray}}
\newcommand{\eea}{\end{eqnarray}}

\begin{document}

\preprint{AIP/123-QED}

\title{Self Consistent Field Theory of Virus Assembly}

\author{Siyu Li}
\affiliation{Department of Physics and Astronomy,
   University of California, Riverside, California 92521, USA}
\author{Henri Orland}
\affiliation{Institut de Physique Th$\acute{e}$orique, CEA-Saclay, CEA, F-91191 Gif-sur-Yvette, France}
\affiliation{Beijing Computational Science Research Center, No.10 East Xibeiwang Road, Haidan District, Beijing 100193, China }
\author{Roya Zandi}
\affiliation{Department of Physics and Astronomy,
University of California, Riverside, California 92521, USA}
\email{roya.zandi@ucr.edu}

\date{\today}

\begin{abstract}
The Ground State Dominance Approximation(GSDA) has been extensively used to study the assembly of viral shells. In this work we employ the self-consistent field theory (SCFT) to investigate the adsorption of RNA onto positively charged spherical viral shells and examine the conditions when GSDA does not apply and SCFT has to be used to obtain a reliable solution. We find that there are two regimes in which GSDA does work. First, when the genomic RNA length is long enough compared to the capsid radius, and second, when the interaction between the genome and capsid is so strong that the genome is basically localized next to the wall. We find that for the case in which RNA is more or less distributed uniformly in the shell, regardless of the length of RNA, GSDA is not a good approximation. We observe that as the polymer-shell interaction becomes stronger, the energy gap between the ground state and first excited state increases and thus GSDA becomes a better approximation. We also present our results corresponding to the genome persistence length obtained through the tangent-tangent correlation length and show that it is zero in case of GSDA but is equal to the inverse of the energy gap when using SCFT.  
\end{abstract}

\maketitle


\section{\label{intro}Introduction}

Viruses have evolved to optimize the feat of genome packaging inside a nano-shell called the capsid, built from several copies of either one or a few different types of proteins.  Quite remarkably, under many circumstances the capsid proteins of single-stranded RNA viruses can assemble spontaneously\cite{elife,Cornelissen2007,Ren2006,Bogdan,Anze2,Zlotnick,Sun2007,Nature2016,Luque:2010a} around the cognate and non-cognate RNAs and other negatively charged cargos\cite{Sun2007,Kusters2015,Zandi2016,Bruinsma2016,Stockley2013}. It is widely accepted that the electrostatic interaction is the main driving force for the assembly\cite{Cornelissen2007,Ren2006,Bogdan,Anze2,Zlotnick,Hsiang-Ku,Venky2016,Shklovskii} and it is this feature that has made viruses ideal for various bio-nanotechnological applications including gene therapy and drug delivery.  

Despite their great interest in biological and industrial applications, the physical factors contributing to the efficient assembly and stability of virus particles are not well understood \cite{Wagner2015956,adsorption2015}.  The difficulty emerges from the considerable number of variables in the system including the genome charge density, the persistence length, the surface geometry and the charge density of surface charges.  The adsorption of genome to the inner wall of capsid, the interplay between long-range electrostatic and short-range excluded volume interactions and the issue of chain connectivity make the understanding of the problem quite challenging.  The presence of salt makes the adsorption process even more complicated. The salt ions can screen the electrostatic interaction between the charges and modify the persistence length of the genome leading to a change in the profile of the genome in the capsid. 

Because of the difficulties noted above, in all previous studies on the encapsidation of viral genome by capsid proteins, the ground state dominance approximation, in which only the lowest energy eigenstate of the system is considered, has been exclusively used\cite{Vanderschoot,Vanderschoot2009,Gonca2014,Gonca2016,Li2017,Siber2008,SiberZandi2010,GoncaPRL2017}. In this paper, we investigate the validity of GSDA in different regimes as a function of salt concentration, genome charge density and surface charge density. Note that viral RNA is relatively long compared to the capsid inner radius.  For example for many plant viruses, RNA is about 3000 nucleotides while the inner capsid radius is around 10 $nm$\cite{Comas}. While it is well-known that GSDA works well for long chains\cite{deGennes1979}, in many recent virus assembly experiments short pieces of RNA have been systematically employed, to study the impact of genome length on the virus stability and formation\cite{Tuli2017}.  Thus the time is ripe to explore the conditions under which GSDA does not apply and self consistent field theory has to be solved to obtain the correct solution.  Comparing the solutions of SCFT and GSDA shows that GSDA is less accurate when the interaction of genome with the capsid wall is weak even if the genome is long.

The paper is organized as follows. In the next section, we introduce the model and all the relevant equations. In Section III, we present our results and discuss the impact on the genome profile of the capsid charge density, salt concentration and polymer length and charge density in Section IV . Finally, in Section IV, we present our conclusion and summarize our findings.




\section{\label{theory}Theory}
In order to calculate the free energy of a virus particle in a salt solution, we model the capsid as a positively charged shell, in which a negatively charged flexible linear polymer (genomic RNA) is confined. Defining by $N$ the number of monomers, $N_+$ the number of salt cations and $N_{-}$ the number of salt anions, the partition function of the system can be written as

\begin{align}\label{eq:Z}
Z \hspace{0.2cm}=\hspace{0.2cm}
&\sum_{i}^{N_+}\sum_{i}^{N_-}\frac1{N_+!}\frac1{N_-!} e^{\beta \mu N_+} e^{\beta \mu N_-} \nonumber\\
&\int \mathcal{D}r_{i}^{+}\mathcal{D}r_{i}^{-}\mathcal{D}r_s 
\exp\left\{- \frac3{2a^2} \int_0^N ds \dot r_s^2 \right. \nonumber \\
&\hspace{0.75cm}-\frac1 2\int dr dr'\hat{\rho}_m(r) u(r-r') \hat{\rho}_m(r') \nonumber \\
&\hspace{1.5cm}\left.-\frac\beta 2 \int drdr' \hat{\rho}_c(r)\upsilon_c(r-r')\hat{\rho}_c(r') \right\} 
\end{align}
where $a$ is the Kuhn length of the monomers. We assume that the salt is monovalent (charge $e$ per ion), and the charge per monomer is $\tau$. The monomer density $\hat{\rho}_m(r)$ and the charge density $\hat{\rho}_c(r)$ are given by
\begin{align} 
&\hat{\rho}_m(r)=\int_0^N ds \delta(r-r_s) \label{rhom}\\
&\hat{\rho}_c(r)=\rho_0(r) +\tau \int_0^N\delta(r-r_s)ds \nonumber\\
&\hspace{2.5cm}+e \left( \sum_i^{N^+}\delta(r-r_i^+)- \sum_i^{N^-} \delta(r-r_i^-) \right) \label{rhoc}
\end{align}
where $\rho_0(r)$ denotes the charge density of the viral shell.  In Eq.~(\ref{eq:Z}), the term $u(r)= u_0 \delta(r)$ represents Edwards's excluded volume interaction, and $v_c(r) = 1/4 \pi \epsilon r$ is the Coulomb interaction between the charges, where $\epsilon$ is the dielectric permitivity of the solvent.

\subsection{Self Consistent Field Theory}
To obtain the genome profile inside the virus capsid, we use Self-Consistent Field Theory (SCFT \cite{Edwards:65}) and the grand canonical ensemble for the salt ions with their fugacity $\lambda$ corresponding to the concentration of salt ions in the bulk. Performing two Hubbard-Stratonovich transformations and introducing the excluded volume field $w(r)$ and the electrostatic interaction field $\phi$ (see Supplementary Material), Eq.~\ref{eq:Z} simplifies to
\begin{align*}\label{HS_Z}
&\mathcal{Z}=
\int \mathcal{D}w(r)\mathcal{D}\phi(r) \\
&~~e^{ \log Q-\int dr \{ \frac1{2u_0} w^2(r)+\frac{\beta\epsilon}{2}\left({\nabla \phi(r)}\right)^2 -2\lambda \cosh(i\beta e \phi(r)) + i\beta\rho_0(r)\phi(r) \} }  
\end{align*}
where $Q$ denotes the partition function for a single chain 

\begin{equation}
\label{single}
Q=\int\mathcal{D}r_s e^{-\frac3{2a^2} \int_0^N ds \dot r_s^2 -i \int dr \hat{\rho}_m(r)[w(r)+\beta\tau \phi(r)]}.
\end{equation}
The Self-Consistent Field Theory equations are obtained by performing the saddle-point approximation on the two integration fields $w$ and $\phi$, see Supplementary Material. The equations are
\begin{eqnarray}
w(r) &=& u_0 \rho_m(r) \\
-\epsilon \nabla ^2 \phi &=& -2 \lambda e \sinh \left(\beta e \phi(r)\right) + \rho_0(r) + \beta \tau \rho_m(r) \label{SPM}
\end{eqnarray}
where
\begin{equation}
\label{density}
\rho_m(r) =\int_0^N ds~q(r,N-s)q(r,s)
\end{equation} 
is the monomer concentration at point $r$. Equation \ref{SPM} is the Poisson-Boltzmann equation for the charged monomers-salt ions system \cite{Borukhov:98a}.
 
 In Eq.~\ref{density}, we have introduced the propagator $q(r,s)$, which is proportional to the probability for a chain of length $s$ to start at any point in the viral shell and to end at point $r$ \cite{fredrickson}. It satisfies the SCFT (diffusion) equation \cite{Doi:86},
 \begin{eqnarray}
 \frac{\partial q(r,s)}{\partial s}&=&\frac{a^2}{6}\nabla^2 q(r,s) - V(r)q(r,s) \label{eq:SCFT} \\
 V(r) &=& w(r) + \beta \tau \phi(r) \label{V}
\end{eqnarray} 
with the following boundary condition
\begin{eqnarray} \label{eq:BC}
&&q(r,0) = \frac{1}{\sqrt{Q}} 
\eea
for $r$ anywhere in the virus shell. The single chain partition function ${Q}$ is given in Eq.~\ref{single} and is determined through the normalization condition on $q(r,s)$
\be \label{eq:Norm}
\int_0^N dr q(r,N-s)q(r,s) = 1 {\text {\rm \ for\ any}} \ s 
\ee
Note that the SCFT Eq.~\ref{eq:SCFT} can also be written as an imaginary time Schr\"odinger equation in the form
\be
\frac{\partial q(r,s)}{\partial s} = -H q(r,s)
\ee
with the Hamiltonian $H$ given by
\be
H = - \frac{a^2}{6}\nabla^2 +V(r)
\ee
Once we obtain the propagator $q$ then we can calculate the chain persistence length or stiffness as explained in the next section.
\subsection{Persistence Length}
Polymers may have some bending rigidity or stiffness, due either to their intrinsic mechanical structure or to the Coulombic interaction between charged monomers, which has a tendency to rigidify the chain. This stiffness results in a strong correlation between the orientation of successive monomers. Eventually, at large separations, the directions of monomers become uncorrelated. The persistence length of a polymer is the correlation length of the tangents to the chain \cite{Abels2005,Doi:86}. It is the typical distance over which the orientation of monomers becomes uncorrelated. The chain can be viewed as a set of independent fragments of length equal to their persistence length.

In order to compute the persistence length, we calculate the correlation function of tangents to the chain
\be
C(s,s') = \langle \dot r(s) \dot r(s') \rangle.
\ee
 We show in Supplementary Material that within the SCFT, this correlation function can be expressed as
 \bea
 \label{persistence}
 C(s,s') &=& \frac{a^4}{9} \int dr dr' \left( \frac{\partial}{\partial r} q(r,N-s) \right) \left( \frac{\partial}{\partial r'} q(r',s') \right) \nonumber \\
  && \times \langle r | e^{-(s-s')H} | r' \rangle
\eea
where we assumed that $s > s'$. In this equation, for brevity we have used the standard quantum mechanical representation for the matrix elements of the evolution operator, see for example Eq.~S5,~S9,~S28 in Supplementary Material. 

For large separation $s-s' \gg 1$, this function behaves as
\be
C(s,s') \approx e^{-(s-s')/l_p}
\ee
where by the above definition, $l_p$ is the persistence length of the chain.

\subsection{Ground State Dominance Approximation}
The set of non-linear partial differential equations given in Eqs.~\ref{SPM},~\ref{eq:SCFT} are very tedious to solve. In the case of a confined chain, or more generally for a system with a gap in the energy spectrum of the Hamiltonian $H$, it is convenient to use the so-called Ground State Dominance Approximation as noted in the introduction. This approximation consists of expanding the propagator $q$ (Eq.~\ref{eq:SCFT}) in terms of the eigenfunctions of the Hamiltonian $H$. We thus write
\be
\label{expansion}
q(r,s) = \sum_{k=0}^{\infty} e^{-E_k s} q_k \psi_k(r)
\ee
where $\{E_k, \psi_k(r), k = 0, 1, 2, \ldots \}$ are the set of normalized eigenvalues and eigenstates of $H$, respectively,
\bea
H \psi_k(r) &=& E_k \psi_k(r) \nonumber \\ 
\int dr \ \psi_k^2(r) &=& 1.
\eea
Using the boundary condition Eq.~\ref{eq:BC}, we find
\bea \label{eq:qk}
q_k &=& \frac{1}{\sqrt {Q}}\int dr  \psi_k(r)
\eea
with
\bea  \label{eq:Q}
{Q} &=& \sum_{k=0}^{\infty} e^{-NE_k} \left( \int dr  \psi_k(r) \right)^2
\eea
We assume that the eigenvalues are ordered as $E_0 < E_1< \ldots< E_k < \ldots$.
When the energy gap between the ground state $E_0$ and the first excited state $E_1$ is large, the ground state 
dominates the expansion Eq.~\ref{expansion} and we may write
\be
q(r,s) = e^{-E_0 s} \left( q_0 \psi_0(r) + e^{-s \Delta} R(r,s) \right)
\ee
where $\Delta = E_1 - E_0 $ is the energy gap, and the function $R(r,s)$ is the remainder of the expansion. When $s\Delta \gg 1$, the second term above becomes exponentially negligible, and we may write
\be
\label{GSDA}
q(r,s) = e^{-E_0 s} q_0 \psi_0(r) 
\ee
and then Eqs.~\ref{eq:Q}, \ref{eq:qk} and \ref{density} become respectively equal to 

\bea
&Q=e^{-N {E}_0} \left( \int dr  \psi_0(r) \right)^2\\ \label{eq:q0}
& q_0 = e^{NE_0/2} \\ 
&\rho_m(r) = N \psi_0^2(r)
\eea

The Poisson-Boltzmann (Eq.~\ref{SPM}) and diffusion (Eq.~\ref{eq:SCFT}) equations then  become
\bea \label{GSDA}
&-\epsilon\nabla^2\phi=-2\lambda e \sinh(\beta\phi)+N\tau \psi_0(r)^2+\rho_0 \nonumber\\
&-\frac{a^2}{6}\nabla^2 \psi_0(r) + N u_0 \psi_0(r)^3 + \beta\tau \phi(r)\psi_0(r) = E_0 \psi_0(r) \nonumber\\
\eea
and the energy $E_0$ is determined so that $\psi_0$ is normalized as
\be
\int dr \ \psi_0^2(r) =1
\ee

Similarly, we can compute the correlation function Eq.~\ref{persistence} within the GSDA. Using Eq.~\ref{eq:q0} and the fact that
\be
\langle r | e^{-(s-s')H} | r' \rangle = e^{-(s-s')E_0} \psi_0(r) \psi_0(r')
\ee
in GSD,
we obtain
\bea
C(s,s') &=& \frac{a^4}{9} \left( \int dr \psi_0(r) \frac{\partial \psi_0}{\partial r} \right)^2 \nonumber \\
&\equiv& 0
\eea
since the integral is identically 0. We conclude that in the GSDA, the persistence length vanishes. 
In order to have a non-vanishing persistence length, we need to include more than the ground state in the eigenstate expansion of all quantities. Including the next leading order term (first excited state with energy $E_1$ and wave function $\psi_1$), we obtain (see Supplementary Material)
\be
C(s,s') \approx A_1 e^{-|s-s'|\Delta} + A_2  e^{-(N-|s-s'|)\Delta}
\ee
which shows that the persistence length is the inverse of the gap
\be
\label{GSDlp}
l_p = \frac{1}{\Delta}
\ee
The persistence length can be computed using the GSDA as it follows: having solved the GSD Eqs.~\ref{GSDA}, we know $E_0$, $\psi_0(r)$ and $\phi(r)$ from which we can calculate $q(r,s)$ and the Hamiltonian $H$. We can then compute the first excited state of $H$ with energy $E_1$, and then the persistence length $l_p$ from Eq.~\ref{GSDlp}.

\section{\label{result}Results}

Due to the complexity of the problem, we numerically solve the non-linear coupled equations given in Eqs.~\ref{SPM} and \ref{eq:SCFT}.  We consider two different cases for the interaction of genome with the capsid. First we study the adsorption of the chain to the capsid inner wall in the absence of the electrostatic interactions, as explained in section III A below. This way we decrease the number of parameters in the system, which helps us to gain some insights before solving the full problem. Then in section III B, we assume that both the capsid and chain are charged in salt solution. 

\subsection{\label{SP}Confined RNA with Adsorption on Capsid}
We consider the confined RNA adsorbed on the capsid wall with no electrostatic interaction present. Thus, the external field ($V_{ext}$) in Eq.~\ref{eq:SCFT} contains only the excluded volume interaction between monomers($u_0$), with an extra attraction from the capsid $\gamma_s$. To solve the diffusion Eq.~\ref{eq:SCFT} with this surface term is not trivial, the strategy we introduce therefore is the effective boundary condition \cite{Hone}:
\begin{equation}
\bigg[\frac{\partial}{\partial r}q(r,s)-\kappa q(r,s) \bigg]_{r=R}=0
\label{robin}
\end{equation}
where $\kappa^{-1}$ is the extrapolation length and is proportional to the inverse of $\gamma_s$.\\
We employ both SCFT and GSDA to solve the problem of a chain confined in an adsorbing spherical shell. To obtain the exact solutions for SCFT, we solve Eqs.~\ref{SPM} and \ref{eq:SCFT} recursively until conditions in Eqs.~\ref{eq:BC}, \ref{eq:Norm} and \ref{robin} are satisfied. We employ Crank-Nicolson scheme and Broyden method\cite{Xingkun,Wang2004} to solve the relevant equations. For the approximative solutions of GSD, we operate on the coupled nonlinear equations (Eq.~\ref{GSDA}) with finite element method and deal with the convergence issue using Newton method. \\
\begin{figure}[tb]
    \centering
    \includegraphics[width=\linewidth]{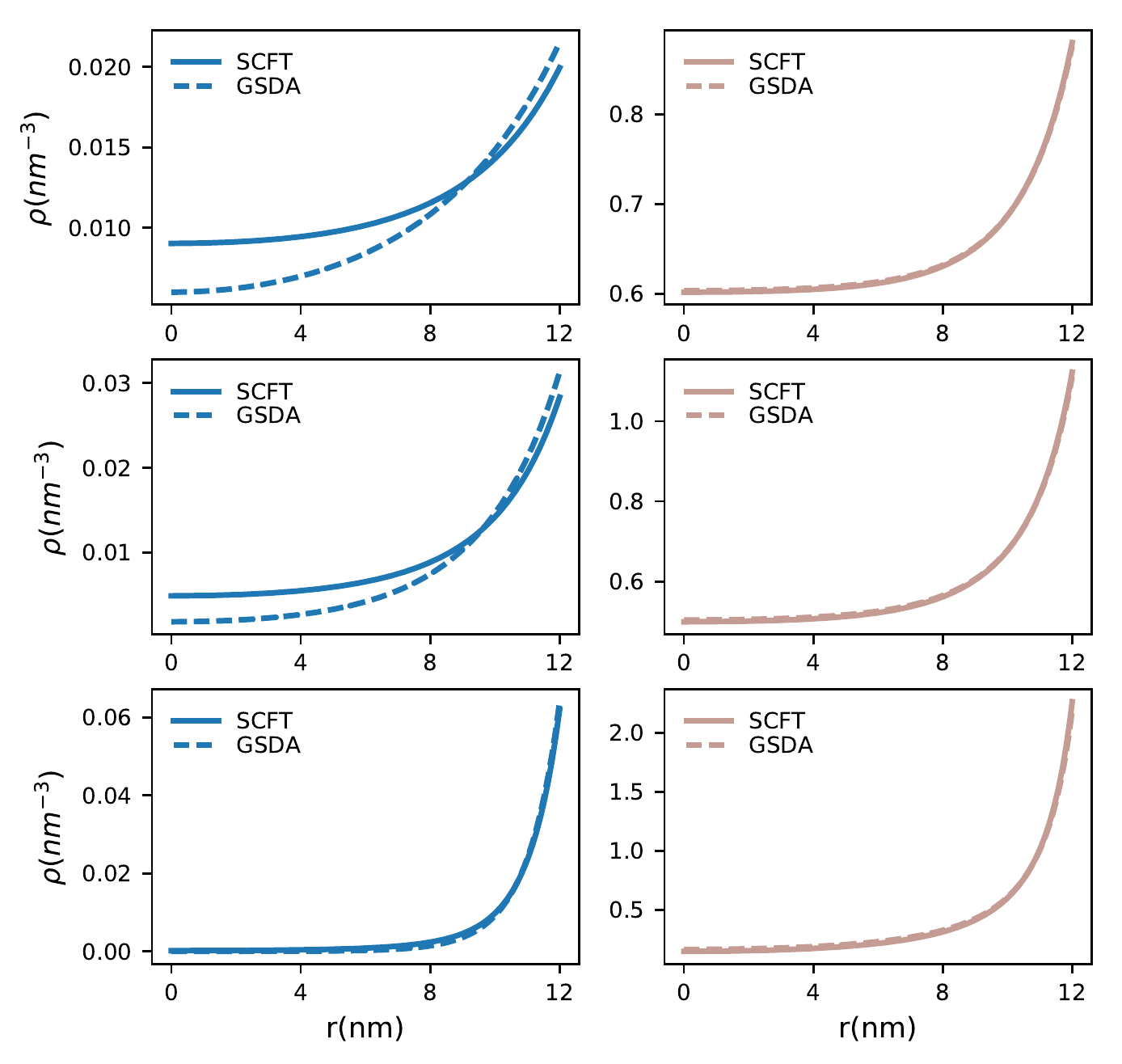}
    \caption{\footnotesize Confined RNA density profile vs $r$ the distance from the capsid center for various extrapolation lengths, $\kappa^{-1}$=\SIlist{10.0;5.0;2.0}{nm} for top to the bottom of the figure.  The total monomer number is N=100(left), N=5000(right).}
\label{robinfig}
\end{figure}
The results of our calculations are presented in Fig.~\ref{robinfig}, which shows the confined RNA density profile as a function of $r$, the distance from the shell center, for various extrapolation length ($\kappa^{-1}$). The goal is to compare our findings obtained through GSDA and SCFT methods for both short and long RNAs. The dashed lines in Fig.~\ref{robinfig} are obtained using GSDA while solid lines are calculated based on the SCFT method. As illustrated in the figure, GSD only makes a good approximation for long chains and/or short extrapolation lengths (strong adsorption regime or large $\kappa$). With short RNA or long extrapolation length(weak adsorption regime), GSDA profile deviates considerably from self-consistent profile. As illustrated in Fig.~\ref{robinfig}, for $N=5000$ regardless of the strength of interaction $\kappa^{-1}$, the solutions of GSDA and SCFT match almost perfectly and completely cover each other.  However, the agreement between the two methods becomes less for $N=100$ and small values of $\kappa$. In the next section, we investigate the impact of electrostatic interaction on the profile of RNA inside the capsid.

\subsection{\label{SPE}Confined RNA with electrostatic interaction}
Since RNA acts like a negatively charged polyelectrolyte in solution, we need to take into consideration the electrostatic interactions term $\beta\tau\phi(r)$ given in Eq.~\ref{eq:SCFT}. We assume that positive charges on the capsid are uniformly distributed. The coulombic interaction does usually overwhelm other forces responsible for the adsorption of chain to the wall, so instead of applying Robin boundary condition (Eq.~\ref{robin}) as in Sect.~\ref{SP}, we use Dirichlet boundary condition ($q(R,s)=0$) for monomer density by assuming the $V_{ext}$ is infinity beyond the capsid wall. The physical basis for this assumption is that RNA monomer has stiffness, and the excluded volume interaction between the capsid wall and the RNA is such that the density of RNA could never sit at the wall. \\
\begin{figure}[H]
   \centering
    \includegraphics[width=\linewidth]{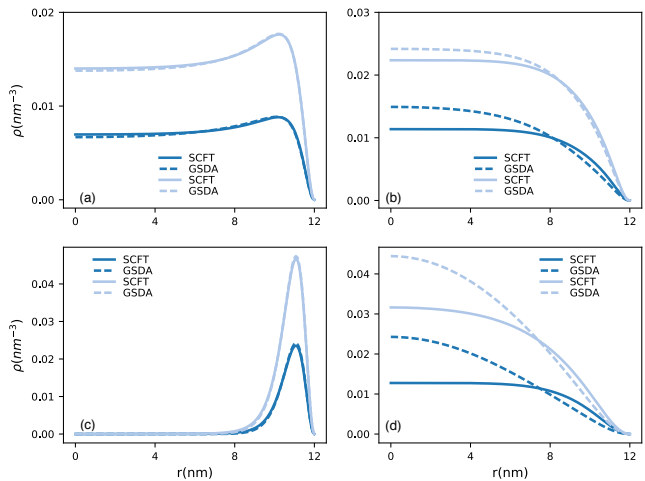}  
    \caption{\footnotesize Confined RNA concentration profiles with various RNA length N=50(darker), N=100(lighter) under SCFT calculation(solid lines) and GSD approximation(dashed lines) with (a) linear chain charge density $\tau$=-1.0\si{\ec}, capsid surface charge density $\sigma$=0.8\si{\ec.nm^{-2}} and salt concentration $\lambda$=500mM; (b)$\tau$=-1.0\si{\ec}, $\sigma$=0.4\si{\ec.nm^{-2}} $\lambda$=500mM; (c)$\tau$=-1.0\si{\ec}, $\sigma$=0.4\si{\ec.nm^{-2}}, $\lambda$=100mM. (d)$\tau$=-0.1\si{\ec}, $\sigma$=0.4\si{\ec.nm^{-2}}, $\lambda$=100mM; Other parameters used are kuhn length $a$=1\si{nm}, excluded volume $u_0$=0.05\si{nm^3}, capsid radius R=12\si{\nm}.}
   \label{varcNsig}
\end{figure}

We then solve Eqs.~\ref{SPM} and \ref{eq:SCFT} to obtain the RNA density through both GSDA and SCFT methods.  The genome concentration profiles are shown in Fig.~\ref{varcNsig} for various RNA length(total monomer number), capsid charge density, chain charge density and salt concentrations. As expected,  there is alway a perfect match between GSDA and SCFT for longer RNAs (large N), while for short RNAs (small N), the energy gap becomes considerable and important, with ground state less dominant in the whole expansion series (Eq.~\ref{expansion}) and GSD approximation becomes less valid. 

We also find that the stronger the electrostatic interaction due to the higher capsid surface charge density or genome linear charge density, the better GSDA and SCFT results agree with each other. Fig.~\ref{varcNsig}a shows that regardless of length of genome, at high surface charge density, GSDA and SCFT give the same results. Note, as we decrease the surface charge density, their difference becomes noticeable, as illustrated in Fig.~\ref{varcNsig}b. However, with lower salt concentration for the same surface charge density as in Fig.~\ref{varcNsig}b, the difference between the two methods once again becomes negligible, Fig.~\ref{varcNsig}c. Quite interestingly as we decrease the chain linear charge density even at low salt, we find again that the agreement between the two models becomes detectable, Fig.~\ref{varcNsig}d.  


All results presented above show that GSDA is less valid when genome localizes close to the center. To this end, we investigate this transition point where the wall attraction becomes so weak that depletion shows up, corresponding to the disappearance of the genome peaks in graphs of Figs.~\ref{varcNsig}a and b and also \ref{profile500}a and ~\ref{profile500}b below.
We calculate the excess genome at the wall by integrating the genome peak area, which is proportional to adsorbed monomers. Then we investigate the impact of the salt concentration and surface charge density on the adsoprtion-depletion transition. The resulting phase diagram is illustrated in Fig.~\ref{phasediagram}.  The white shade in the figure corresponds to the maximum adsorption. As the color gets darker, less genome is adsorbed to the wall. In the darkest region there is no adsorption. The line separating the darkest region indicates the onset of the depletion transition. 

Figure~\ref{profile500} describes the genome profile details for two different cases.  For a fixed salt concentration but varying surface charge density ($\sigma=\SIrange{0}{0.4}{\ec.nm^{-2}}$) we observe that the peak next to the wall slowly disappears as the capsid charge density decreases and most of the genome becomes localized at the center, Fig.~\ref{profile500}a.  Similar behavior is displayed in Fig.~\ref{profile500}b for fixed surface charge density but various salt concentrations. Figs. ~\ref{profile500}a and ~\ref{profile500}b together tell us that the higher salt concentration, or the lower surface density charge, causes genome to stay away from the capsid wall and to localize toward the center, constructing the region where GSDA is not valid any more.

\begin{figure}[tbh]
  \centering
  \includegraphics[width=0.8\linewidth]{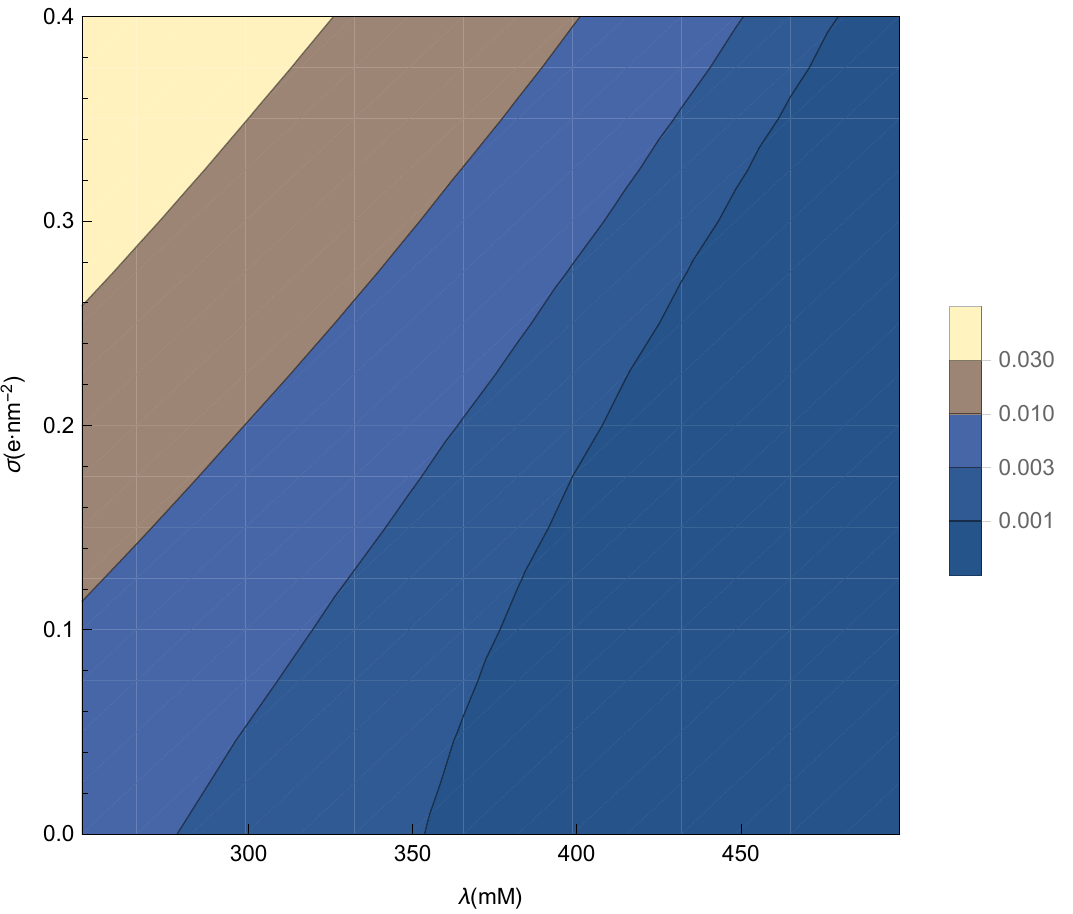}
  \caption{\footnotesize Genome excess phase diagram with respect to salt concentration and capsid surface charge density.  The white shade corresponds to the region with the maximum genome density and black to the depletion regime next to the wall. Other parameters used are $N=500$, $a$=1\si{nm}, $u_0$=0.05\si{nm^3}, R=12\si{\nm}.}
  \label{phasediagram}
\end{figure}
\begin{figure}
  \centering
   \subfloat[]{\includegraphics[width=0.5\linewidth]{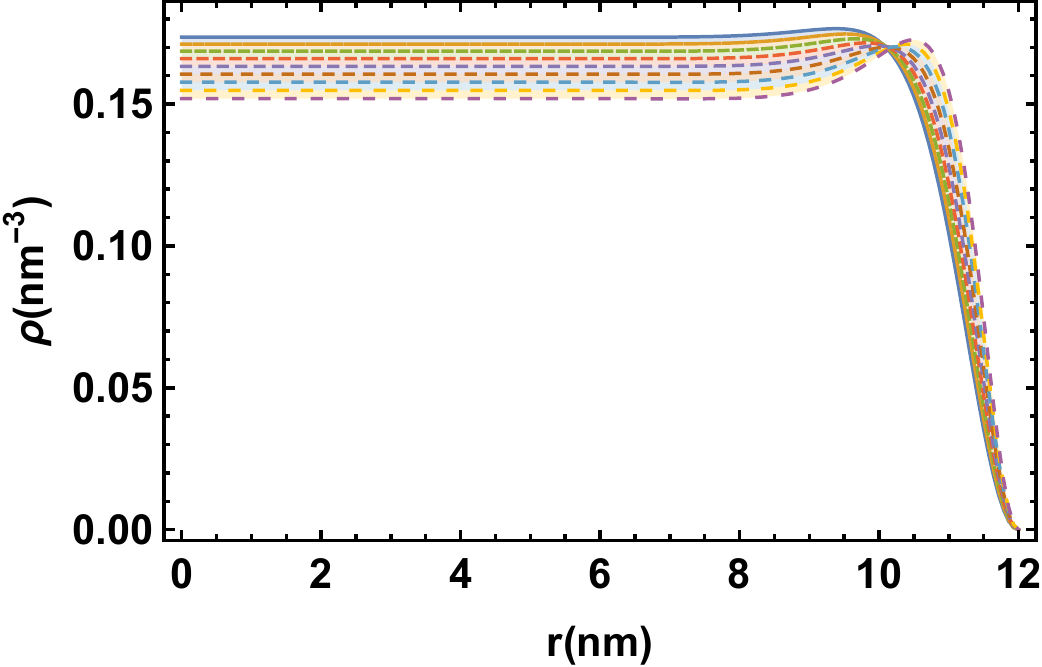}}
   \subfloat[]{\includegraphics[width=0.5\linewidth]{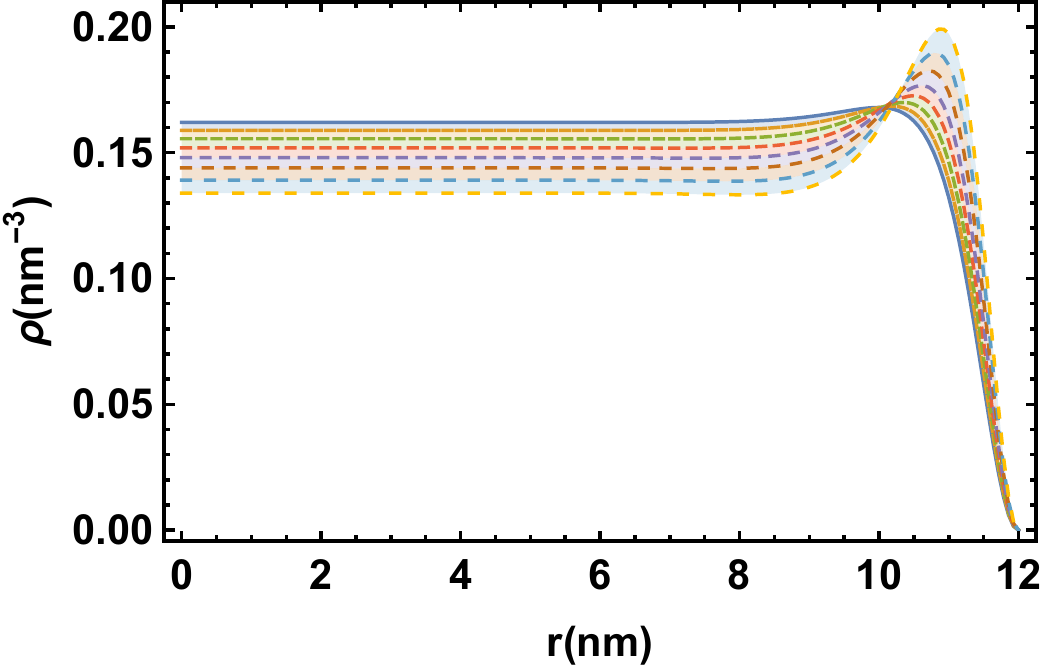}}
   \caption{\footnotesize Genome density profile for $N=1000$ and (a) various surface charge density (\SIrange{0}{0.4}{\ec.nm^{-2}}) with salt concentration $\lambda$=400\si{mM}; (b) various salt concentration(\SIrange{250}{500}{\milli\molar}) with fixed surface charge $\sigma=0.4\si{\ec.nm^{-2}}$. Other parameters correspond to kuhn length $a$=1\si{nm}, excluded volume $u_0$=0.05\si{nm^3}, capsid radius R=12\si{\nm}.}
   \label{profile500}
\end{figure}

\section{\label{discussion}Discussion and Summary}
The results of previous sections show that the GSDA validity depends on the genome localization: when the genome is absorbed on the wall, GSDA works perfectly, however when the adsorption becomes weaker and the genome starts moving to the center, GSDA stops being reliable. Fig.~\ref{varcNsig} illustrates this statement, where perfect match between GSDA and SCFT is obtained in lower salt concentration and higher surface charge (localized genome); significant deviation appears at higher salt concentration and lower surface charge in which case the genome is delocalized. The same effect is observed for the linear charge density of short genomes.

For longer genome with 500 monomers or more, the difference is almost undetectable. Quite interestingly, the effect of the electrostatic interaction range and strength, salt concentration and surface charge density in Sect.~III B is similar to that of the extrapolation length in Sect.~III A.  While low salt concentration(longer Debye length, strong attraction) and high surface charge correspond to larger $\kappa$, high salt concentration (short Debye length, weak attraction) and low surface charge on the contrary correspond to small $\kappa$ in which case the GSDA does not work well as illustrated in Fig.~\ref{robinfig}. 

Another important difference arising from using GSDA and SCFT approaches corresponds to the tangent-tangent correlation function or persistence length of the polymer.  While the persistence length obtained through GSDA is zero, the persistence length calculated using SCFT is inversely proportional to the energy gap between the ground state and the first excited state, Eq.~\ref{GSDlp}. The vanishing persistence length in GSDA is due to the fact that the chain constraint or connectivity is absent, and all monomers are independent. 
In the case of SCFT, the persistence length increases with the length of genome until it saturates to a finite value. Then indeed, as $N$ increases, $l_p \ll N$, explaining again why GSDA becomes more and more valid as the length of the genome increases.

While the persistence length corresponds to the stiffness of the polymer, there is another important length scale in the problem but it is associated with the adsorption of polymer on the inner shell of the capsid. The adsorption of polymers to flat surfaces have been thoroughly studied, but the adsorption to spherical shells is less understood \cite{Joanny:99a,Pincus:84a,Ji:87a,Eisenriegler:77a}. In case of flat surfaces, the Edward's correlation length determines the distance from the wall over which the adsorption layer decays. It goes as $\xi \sim 1/\sqrt{u_0\phi_B}$, with $u_0$ the strength of the excluded volume and $\phi_B$ the bulk polymer density.  

The situation studied in this paper is more complex due to the confinement of the polymer inside a spherical capsid in the presence of electrostatics. Quite interestingly,  Fig.~\ref{profile500}(a) and (b) show there is a point around $r=10$ where all the curves cross. According to the figure, the location of the crossing point does not depend on the salt concentration and capsid charge density.  Since the capsid is a closed shell, we cannot define the bulk density in this problem. However, $\phi_B$ is related to the number of monomers in the capsid.  Figures~\ref{profile100}(a) and (b) illustrate the genome profiles for the same parameters as in Figs.~\ref{profile500}(a) and (b) respectively but using a shorter genome length. The genome length is $N=100$ and $N=1000$ in Figs.~\ref{profile100} and ~\ref{profile500}, respectively.  As illustrated in Figs.~\ref{profile100} all the plots again meet at a particular point but the position of the crossing point is moved compared to Fig.~\ref{profile500}. It is interesting that despite different capsid charge density and salt concentration, all curves again meet at a unique single distance from the wall. 

We also checked the position of the crossing point as a function of the excluded volume interaction expressed through the Edward's correlation length $\xi \sim 1/\sqrt{u_0\phi_B}$.  Our numerical results did not show any dependence of the crossing point on the strength of the excluded volume interaction. This is probably due to the fact that $\phi_B$ in this problem is not really the bulk density and depends on the excluded volume interaction and might cancel the impact of the excluded volume interaction. Although we cannot provide a closed form formula for the Edward's correlation length, it is interesting that all points meet at one single point and this point is independent of the capsid charge density, salt concentration and the polymer excluded volume interaction but depends on the length of genome.
\begin{figure}[H]
\centering
   \subfloat[]{\includegraphics[width=0.5\linewidth]{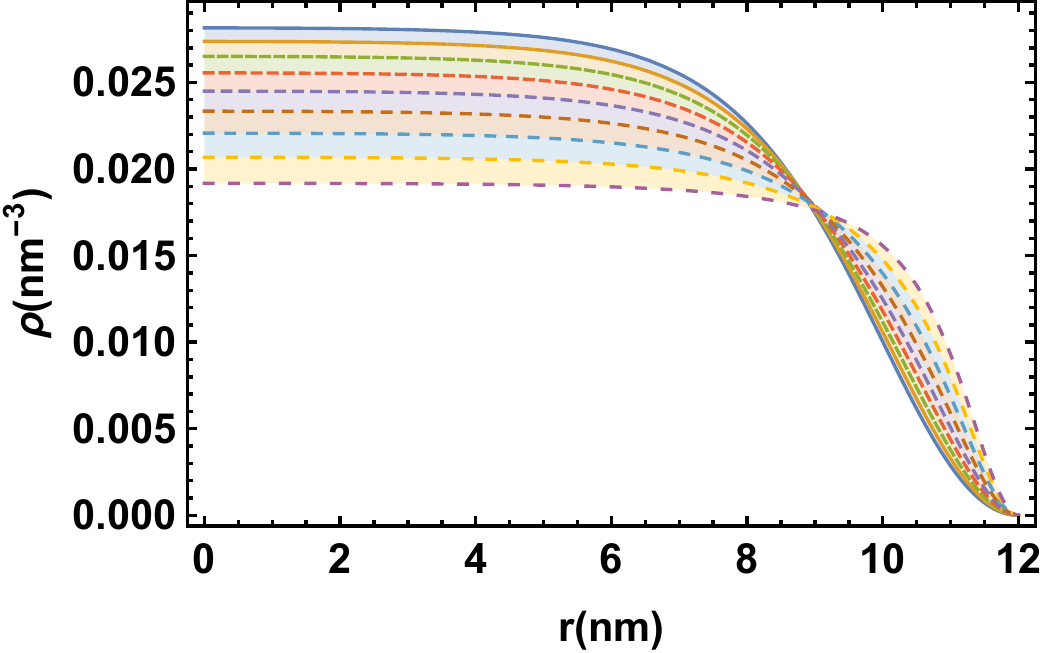}}
   \subfloat[]{\includegraphics[width=0.5\linewidth]{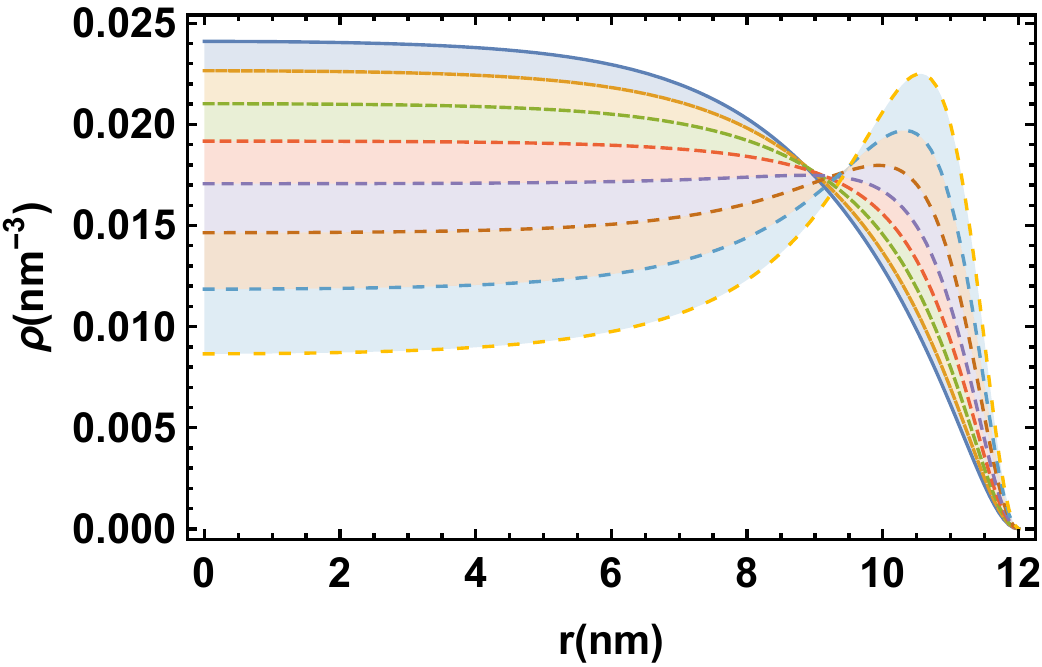}}
   \caption{\footnotesize Genome density profile for $N=100$ and (a) various surface charge density (\SIrange{0}{0.4}{\ec.nm^{-2}}) with salt concentration $\lambda$=400\si{mM}; (b) various salt concentration(\SIrange{250}{500}{\milli\molar}) with fixed surface charge $\sigma=0.4\si{\ec.nm^{-2}}$. Other parameters correspond to $a$=1\si{nm}, $u_0$=0.05\si{nm^3}, R=12\si{\nm}.}.
   \label{profile100}
\end{figure}

In summary, in this paper we investigated the validity of GSDA for studying the profile of genomes in viral shells because of the extensive usage of GSDA in the literature in describing the process of virus assembly and stability. We found that for small RNA segments employed in recent experiments or for {\it in vitro} assembly studies with mutated capsid proteins carrying lower charge density \cite{Brasch2013,Venky2016,Tuli2017,Maassen2017}, the GSDA deviates from the accurate results obtained through SCFT methods. Otherwise, native RNA viruses are long enough compared to the radius of the capsid and as such GSDA is good enough to explain different experimental observations and there is no need to solve tedious self-consistent equations.  Our results showed that the narrower the region RNA is sitting and the stronger is genome-capsid interaction, the larger the energy gap, and hence the better GSDA works. 

%

\section*{Acknowledgments}
The authors would like to thank Xingkun Man for useful discussions. This work was supported by the National Science Foundation through Grant No. DMR -1719550 (R.Z.).

\bibliography{bibfile}
\end{document}


\preprint{AIP/123-QED}

\title{Self Consistent Field Theory of Virus Assembly\\Supplementary Material}

\author{Siyu Li}
\affiliation{Department of Physics and Astronomy,
   University of California, Riverside, California 92521, USA}
\author{Henri Orland}
\affiliation{Institut de Physique Th$\acute{e}$orique, CEA-Saclay, CEA, F-91191 Gif-sur-Yvette, France}
\affiliation{Beijing Computational Science Research Center, No.10 East Xibeiwang Road, Haidan District, Beijing 100193, China }
\author{Roya Zandi}
\affiliation{Department of Physics and Astronomy,
University of California, Riverside, California 92521, USA}
\email{roya.zandi@ucr.edu}

\date{\today}

\maketitle
\section{Self Consistent Field Theory}
In this section we derive several equations presented in Section II of the paper.  Using a Hubbard-Stratanovich transformation,  the partition function of Eq.~1 reads
\begin{eqnarray} 
\label{Z}
&&\mathcal{Z}=\int   
\mathcal{D}w(r) 
\mathcal{D}\phi(r) 
\text{ }e^{\log Q[iw(r)+i \tau \phi(r)]}\nonumber \\
&&e^{
\int dr \left(
-\frac{1}{2 u_0} w(r)^2
-\frac{\beta \epsilon}{2} (\nabla \phi(r))^2
+2 \lambda \cosh\left(i \beta e \phi(r)\right)- i \beta\rho_0(r) \phi(r) \right)}\nonumber \\
\end{eqnarray}
where $\lambda$ is the salt ion fugacity and
\begin{equation}\label{Q1}
Q=
\int\mathcal{D}r(s)
e^{-\frac3{2a^2} \int_0^N ds \dot r^2
- i\int_0^N ds  \left( w(r(s)) +  \beta \tau \phi(r(s))\right)}
\end{equation}
is the partition function of a single Gaussian chain in the external field $\left(iw(r)+i \beta \tau \phi(r) \right)$.  Now performing the saddle-point approximation on the two fields $w(r)$ and $\phi(r)$, we get from Eq.~\ref{Z} 
\begin{align}
w(r)&=u_0 \frac{\delta \log Q}{\delta w(r)} \label{sdrho1}\\
-\epsilon \nabla^2 \phi(r)&=2 \lambda i e \sinh(i\beta e\phi)-i\rho_0(r)+\frac{\delta \log Q}{\delta \phi(r)} \label{sdphi1}
\end{align}
At this stage, from Eqs.~(\ref{sdrho1}) and (\ref{sdphi1}), we can anticipate that the fields $w(r)$  and $\phi(r)$ are pure imaginary and redefine $iw \rightarrow w$ and $i\phi \rightarrow \phi$.

Using standard quantum mechanical notations, we may write
\begin{equation}
\label{Q}
Q= \int dr dr' \langle r |e^{- N H} |r' \rangle
\end{equation}
where the Hamiltonian $H$ is given by
\begin{equation}
\label{ham}
H = -\frac{a^2}{6} \nabla^2 +( w(r) +  \beta \tau \phi(r))
\end{equation}

From Eqs.~(\ref{sdrho1}) and (\ref{Q1}), we have
\begin{eqnarray}
\frac{\delta \log Q}{\delta w(r)}
&=&\frac{1}{Q}
\int\mathcal{D}r(s) \int_0^N ds \delta(r-r{(s)}) \nonumber\\
&&e^{-\frac3{2a^2} \int_0^N ds \dot r^2
- \int ds \left( w(r(s)) +\beta \tau \phi(r(s))\right)}  \nonumber \\
&=& \rho (r)
\end{eqnarray}
where $\rho(r)$ is the monomer density.

Using quantum mechanical notation, we have
\begin{eqnarray}
\frac{\delta \log Q}{\delta w(r)} 
=\frac{1}{Q}\int_0^N ds \int dr_1 \ dr_2 \langle r_1 | e^{-(N-s)H} |r \rangle \langle r |  e^{- sH} | r_2\rangle \nonumber\\
\end{eqnarray}

Defining the propagator $q(r,s)$ as
\begin{equation}
\label{prop}
q(r,s) = \frac{1}{\sqrt{Q}} \int dr' \langle r |  e^{- sH} | r'\rangle
\end{equation}
we have
\begin{eqnarray}
\label{rho}
\frac{\delta \log Q}{\delta w(r)} &=& \rho(r) \nonumber \\
&=& \int_0^N ds \ q(r,N-s) q(r,s)
\end{eqnarray}

Using the above equations, Eqs.~(\ref{sdrho1}),~(\ref{sdphi1}) become
\begin{align}
w(r)&=u_0 \rho(r) \label{sdrho2}\\
-\epsilon \nabla^2 \phi(r)&=-2 \lambda e \sinh(\beta e\phi)+\rho_0(r)+ \beta \tau \rho(r) \label{sdphi2}
\end{align}
and the propagator $q(r,s)$ satisfies the diffusion equation
\begin{equation}\label{diffusion}
\frac{\partial q(r,s)}{\partial s}=\frac{a^2}{6}\nabla^2 q(r,s) - \left(w(r) + \beta \tau \phi(r)\right)q(r,s)
\end{equation}
which can be recast in the form of the Schr\"odinger equation
$\left (\frac{\partial }{\partial s}+H  \right) q(r,s)=0$ with  the Halmitonian $H$ given by Eq.~\ref{ham}.

Numerically, we solve Eqs.~(\ref{rho}), (\ref{sdrho2}), (\ref{sdphi2}) and (\ref{diffusion}) recursively until convergence, with initial condition $q(r,0)=1/\sqrt{Q}$

\section{Ground State Dominance Approximation}
Expanding the partition function $Q$ as well as the propagator $q$ on the normalized eigenstates $|\psi_n \rangle $ with eigenvalues $E_n$ of the Hamiltonian $H$
\begin{eqnarray}
H |\psi_n \rangle &=& E_n |\psi_n \rangle  \\
\int dr \psi_n^2(r) &=&1
\end{eqnarray}
we have
\begin{eqnarray}
\label{Qexp}
Q&=& \int dr dr' \langle r |e^{- N H} |r' \rangle \nonumber \\
&=& \sum_n\int dr dr \psi_n(r) \psi_n(r') e^{-N E_n} 
\end{eqnarray}

Reducing the expansion (\ref{Qexp}) to the lowest eigenstates of $H$, we see that we can keep only the ground state $\psi_0$ with energy $E_0$, provided the first excited state $\psi_1$ has an eigenvalue $E_1$ such that 
\begin{equation}
N \Delta \ll 1
\end{equation}
where $\Delta = E_1-E_0$ is the gap in the eigenvalues.

This is the so-called {\em Ground State Dominance} approximation (GSDA). It implies
\begin{align} 
\label{GSDAQ}
Q \approx I_0^2 e^{-N E_0}
\end{align}
where 
$I_0=\int dr \psi_0(r)=\sqrt{Q} e^{\frac{N E_0}{2}} $ and
\begin{align} \label{GSDAq}
q(r,s)
&\approx \frac{1}{\sqrt{Q}}e^{-s E_0} \psi_0(r)I_0 \nonumber\\
&=e^{\frac{N E_0}{2}} e^{-s E_0} \psi_0(r)
\end{align}
and the density (\ref{rho}) becomes
\begin{align} 
\label{GSDArho}
\rho(r) =N \psi_0^2(r)
\end{align}

The coupled equations (\ref{sdrho2},~\ref{sdphi2}) can be written as
\begin{align}
-E_0 \psi_0(r)&=\frac{a^2}{6}\nabla^2 \psi_0(r) - \left(u_0 N\psi_0(r)^2 + \beta \tau \phi(r)\right)\psi_0(r)  \\
-\epsilon \nabla^2 \phi(r)&=-2 \lambda e \sinh(\beta e\phi)+\rho_0(r)+ \beta \tau N \psi_0(r)^2 
\end{align}

Using the normalization  $\sqrt{N} \psi_0(r) \rightarrow \psi_0(r)$, the above equations become the familiar Euler-Lagrangian equations\cite{Borukhov1,Gonca2016,Li2017}
\begin{align}
-E_0 \psi_0(r)&=\frac{a^2}{6}\nabla^2 \psi_0(r) - u_0 \psi_0(r)^3 - \beta \tau \phi(r)\psi_0(r) \label{GSDAdiffusion} \\
-\epsilon \nabla^2 \phi(r)&=-2 \lambda e \sinh(\beta e\phi)+\rho_0(r)+ \beta \tau \psi_0(r)^2 \label{GSDAsdphi}
\end{align} 
with the constraint
\begin{equation}\label{N}
N=\int \psi_0(r)^2 dr
\end{equation}

\section{Persistence length}
The persistence length is the correlation length of the tangents of the polymer.
\begin{equation}
C(s,s')=\frac{1}{a^2}\left(<\dot r(s)\cdot \dot{r}(s')>-<\dot{r}(s)>\cdot<\dot{r}(s')>\right)
\end{equation}
where the average is calculated with the Boltzmann weight given by the partition function Eq.~\ref{Z}.
Note that the true correlation length requires to use unit tangent vectors rather than $\dot r$.
With spherical geometry we have $<\dot{r}(s)>=0$, and the persistence length is given by
\begin{eqnarray}
\label{C1}
C(s,s')
&=&\frac{1}{a^2} <\dot r(s)\cdot \dot{r}(s')> \nonumber\\
&=&\frac{1}{a^2} \frac{1}{\mathcal{Z}_0} \int  \mathcal{D}w(r) \mathcal{D}\phi(r) \nonumber \\
&e&^{\int dr \left(
-\frac{1}{2 u_0} w(r)^2
-\frac{\beta \epsilon}{2} (\nabla \phi(r))^2
+2 \lambda \cosh\left(i \beta e \phi(r)\right)- i \beta\rho_0(r) \phi(r) \right)} \nonumber\\
&&\times \int {\cal D} r(s) \dot{r}(s) \dot{r}(s')
e^{-\frac3{2a^2} \int_0^N ds \dot r_s^2
- \int_0^N ds \left(w(r(s)) + \beta \tau \phi(r(s))\right)} \nonumber \\
\end{eqnarray}

Fixing the fields $w(r)$ and $\phi(r)$ to their SCFT value, the above expression (\ref{C1}) simplifies to 
\begin{align}
C&(s,s') \nonumber\\
&=\frac{1}{a^2}\frac1{Q}
\int \mathcal{D}r_s~
\dot{r}(s) \dot{r}(s')
e^{-\frac3{2a^2} \int_0^N ds \dot r_s^2
- \int_0^N ds \left(w(r(s)) + \beta \tau \phi(r(s) )\right)} \nonumber\\ 
&=\frac{1}{a^2}\frac1{Q}
\int dr_N dr_0 dr dr'  \dot{r}(s) \dot{r}(s') \nonumber\\
&<r_N|e^{-(N-s)H}|r> <r|e^{-(s-s')H}|r'><r'|e^{-s' H}|r_0>
\end{align}
where we have assumed $s > s'$.
Using the correspondance principle 
\begin{eqnarray}
p&=&\frac{1}{i}\frac{\partial}{\partial r} \nonumber \\
\dot{r}&=&\frac{1}{i}\frac{a^2}{3}p=-\frac{a^2}{3}\frac{\partial}{\partial r}
\end{eqnarray}
which is a direct property of the Fourier transform, we obtain the persistence length as
\begin{eqnarray}\label{C2}
C(s,s') &=&\frac{1}{a^2 Q}\int dr_N dr_0 dr dr' 
\left(-\frac{a^2}{3}\frac{\partial}{\partial r}\right)\left(-\frac{a^2}{3}\frac{\partial}{\partial r'}\right) \nonumber \\
&\times&<r_N|e^{-(N-s)H}|r]> <r|e^{-(s-s')H}|r'><r'|e^{-s' H}|r_0>  \nonumber\\
&=&\frac{a^2}{9Q}\int dr dr' \int dr_N\left(\frac{\partial}{\partial r}<r_N|e^{-(N-s)H}|r>\right) \nonumber \\
&\times&<r|e^{-(s-s')H}|r'> 
\int dr_0\left(\frac{\partial}{\partial r'}<r'|e^{-s' H}|r_0>\right) \nonumber\\
&=&\frac{a^2}{9}\int dr dr' 
\left(\frac{\partial}{\partial r}q(r,N-s)\right) \nonumber \\
&\times&<r|e^{-(s-s')H}|r'>\left(\frac{\partial}{\partial r'}q(r',s')\right) \nonumber \\
&=&\frac{a^2}{9}
\sum_n\int dr dr' e^{-(s-s')E_n} \nonumber \\
&\times&\left(\psi_n(r)\frac{\partial}{\partial r}q(r,N-s)\right)
\left(\psi_n(r')\frac{\partial}{\partial r'}q(r',s')\right)
\end{eqnarray}
Having solved the SCFT equations, the full Hamiltonian (\ref{ham}) can be diagonalized numerically and 
the above formula allows to compute the full correlation function, in the SCFT approximation.

If we restrict ourselves purely to the GSDA, using Eqs.~(\ref{GSDAQ}) and (\ref{GSDAq}), we obtain
\begin{eqnarray} 
\label{GSDAC}
C(s,s') &=& \frac{a^2}{9}\int dr dr'  \left(\frac{\partial}{\partial r} e^{\frac{N E_0}{2}} e^{-(N-s) E_0} \psi_0(r)\right) \nonumber\\
&\times&e^{-(s-s')E_0}\psi_0(r)\psi_0(r')
\left(\frac{\partial}{\partial r'} e^{\frac{N E_0}{2}} e^{-{s'} E_0} \psi_0(r')\right) \nonumber \\
&=&\frac{a^2}{9} \int dr dr' \left(\frac{\partial}{\partial r}\psi_0(r')\right) \psi_0(r) \nonumber \\
&\times& \psi_0(r')  \left(\frac{\partial}{\partial r'} \psi_0(r')\right) \nonumber\\
&=&\frac{a^2}{9}
\left(
\int dr \psi_0(r)\frac{\partial}{\partial r}\psi_0(r)\right)^2 \nonumber \\
&=&0
\end{eqnarray}
where we have used the fact that 
\[
\int \psi_0(r) \frac{\partial}{\partial r}\psi_0(r)~dr= \frac{1}{2}\int \frac{\partial}{\partial r}\psi_0^2(r)=  0
\]
Therefore, in the GSDA, the persistence length vanishes, and this is due to the fact that  in this case, the chain constraint is absent.

If the system is confined in the viral shell and there is a sizeable gap, rather than staying in the ground state, we can restrict the expansion to the first excited state. Then 
\begin{equation}
<r'|e^{-s H}|r>=e^{-s E_0}\psi_0(r)\psi_0(r')+e^{-s E_1}\psi_1(r)\psi_1(r')
\end{equation}
and Eq~(\ref{prop}) can be written as
\begin{align}
q&(r,s)
=\frac{1}{\sqrt{Q}}\int dr_0 <r|e^{-s H}|r_0> \nonumber\\
&=\frac{1}{\sqrt{Q}}\left(e^{-s E_0}\psi_0(r)\int dr_0\psi_0(r_0) \right.\nonumber\\
&\hspace{3cm}\left.+e^{-s E_1}\psi_1(r)\int dr_0\psi_1(r_0)\right) \nonumber\\
&=\frac{1}{\sqrt{Q}}\left(e^{-s E_0}I_0 \psi_0(r)
+e^{-s E_1}I_1 \psi_1(r)\right) 
\end{align}
where $I_0=\int dr~\psi_0(r)$, and $I_1=\int dr~\psi_1(r)$.
Plugging this result into Eq.~\ref{C2}, the persistence length is therefore
\begin{align}
&C(s,s') =\frac{a^2}{9 Q}
\sum_{n=0}^{1}\int dr_s dr_{s'} e^{-(s-s')E_n} \nonumber\\
&\times \left(\psi_n(r_s)\frac{\partial}{\partial r_s}\left(e^{-(N-s) E_0}I_0 \psi_0(r_s)+e^{-(N-s) E_1}I_1 \psi_1(r_s)\right)\right) \nonumber\\
&\times \left(\psi_n(r_{s'})\frac{\partial}{\partial r_{s'}}
\left(e^{-s' E_0}I_0 \psi_0(r_{s'})+e^{-s' E_1}I_1 \psi_1(r_{s'})\right)\right) \nonumber\\
&=\frac{a^2}{9 Q} 
\left(e^{-(s-s')E_0} e^{-(N-(s-s')) E_1}I_1^2
\left(\int dr\psi_0(r)\frac{\partial}{\partial r}\psi_1(r)\right)^2 \right.
\nonumber\\
&\hspace{.3cm}\left.
+e^{-(s-s')E_1} e^{-(N-(s-s')) E_0}I_0^2
\left(\int dr\psi_1(r)\frac{\partial}{\partial r}\psi_0(r)\right)^2\right) \nonumber\\
&=A
\left(
e^{-(s-s') \Delta}
\left(\int dr\psi_1(r)\frac{\partial}{\partial r}\psi_0(r)\right)^2
\right.
\nonumber\\
&\hspace{.5cm}\left.
+e^{-(N-(s-s'))\Delta}\left(\frac{I_1}{I_0}\right)^2
\left(\int dr\psi_0(r)\frac{\partial}{\partial r}\psi_1(r)\right)^2\right)
\end{align}
where we have used $\int \psi_0(r) \frac{\partial}{\partial r}\psi_0(r)~dr=0$, $A=\frac{a^2}{9}\frac{1}{Q}e^{-N E_0}I_0^2$, and $\Delta=E_1-E_0$ which is the energy gap.
So for $1<<s-s'<<N$, we can read off the persistence length as
\begin{equation}
l_p=\frac{1}{\Delta}
\end{equation}

\bibliography{bibfile}